\documentclass[a4]{scrartcl}
\usepackage{amsmath,mathrsfs,graphicx,slashed,ragged2e}
\usepackage{amssymb}
\usepackage[numbers,sort&compress]{natbib}
\usepackage{hyperref}
\usepackage[colorinlistoftodos,disable]{todonotes}

\newcommand{\sE}{{\mathbb E}}
\newcommand{\vol}{\operatorname{vol}}

\newcommand{\bt}{\boldsymbol{\theta}}

\newcommand{\bg}{\boldsymbol{\gamma}}

\newcommand{\stodo}[1]{\todo[author=Simon,color=blue,inline]{#1}}
\newcommand{\bill}[1]{\todo[author=Bill,color=red,inline]{#1}}

\listoftodos

\title{The Information Geometry of Sensor Configuration}

\author{Simon Williams \\{University of Melbourne} \and Arthur George Suvorov \\RMIT University \and Zengfu Wang \\{Northwestern Polytechnical University}\and Bill Moran\\{University of Melbourne}}

\date{\today}

\begin{document}

\maketitle

\begin{abstract}
In problems of parameter estimation from sensor data, the Fisher Information provides a measure of the performance of the sensor; effectively, in an infinitesimal sense, how much information about the parameters can be obtained from the measurements. From the geometric viewpoint, it is a Riemannian metric on the manifold of parameters of the observed system. In this paper we consider the case of parameterized sensors and answer the question, ``How best to reconfigure a sensor (vary the parameters of the sensor) to optimize the information collected?'' A change in the sensor parameters results in a corresponding change to the metric. We show that the change in information due to  reconfiguration exactly corresponds to the natural metric on the infinite dimensional space of Riemannian metrics on the parameter manifold, restricted to finite-dimensional sub-manifold determined by the sensor parameters. The distance measure on this configuration manifold is shown  to provide  optimal, dynamic sensor reconfiguration based on an information criterion. Geodesics on the configuration manifold are shown to optimize the information gain but only if the change is made at a certain rate. An example of configuring two bearings-only sensors  to optimally locate a target is developed in detail to illustrate the mathematical machinery, with Fast-Marching methods employed to efficiently calculate the geodesics and illustrate the practicality of using this approach.

\end{abstract}
\section{Introduction}

This paper is an attempt to begin the construction of an abstract theory of sensor management, in the hope that it will help to provide both a theoretical underpinning for the solution of practical problems and insights for future work.  A key component of sensor management is the amount of information a sensor in a given configuration can gain from a measurement, and how that information gain changes as the configuration does.  In this vein, it is interesting to observe how information theoretic \emph{surrogates} have been used in  a range of applications as objective functions for sensor management;  see, for instance, \cite{bell1,donoho1,sameh1}.   Our aim here is to abstract from various papers including these, those by Kershaw and Evans \cite{kershaw} as well as others, the mathematical principles required for this theory.  Our approach is set within the mathematical context of differential geometry.

The problem of estimation is expressed in terms of a likelihood; that is, a probability density  $p(x|\theta)$, where $x$ denotes the measurement and $\theta$ the parameter to be estimated. It is well known,\cite{coverthomas} that the Fisher Information  associated with this likelihood provides a measure of information gained from the measurement. 

While the concepts  discussed here are, in some sense, generic and can be applied to any sensor system that has the capability to modify its characteristics, for simplicity and to keep in mind a motivating example, we focus on a particular problem: that of localization of a \emph{target}.  Of course, a sensor system is itself just a (more complex)  aggregate sensor, but it will be convenient, for the particular problems we will discuss, to assume a discrete collection of disparate (at least in terms of location) sensors, that together provide measurements of aspects of the location of the target.   This distributed collection of sensors, each drawing measurements that provide partial information about the location of a target, using known likelihoods, defines a particular \emph{sensor configuration state}. As the individual sensors  move,  they change their sensing characteristics and thereby the collective Fisher Information associated with estimation of target location.  The Fisher information matrix defines a metric, the Fisher-Rao metric,  over the physical space where the target resides \citep{fisher1,fisher2}, or in more generality  a metric over the parameter space in which the estimation process takes place, and this metric is a function of the location of the sensors.   This observation permits  analysis of the information content of the system, as a function of sensor parameters, in the framework of differential geometry (``information geometry") \citep{amari1,amari2}.  A considerable literature is dedicated to the problem of optimizing the configuration so as to maximize information retrieval (see Sec. II. A) \citep{moran121,moran122,optimpaper,bell1}. The mathematical machinery of information geometry has led to advances in several signal processing problems, such as blind source separation \cite{blind}, gravitational wave parameter estimation \citep{app2}, and dimensionality reduction for image retrieval \citep{app3} or shape analysis \citep{app4}.

In sensor management/adaptivity applications, the performance of the sensor configuration (in terms of some function of the Fisher Information) becomes a cost associated with finding the optimal sensor configuration, and tuning the metric by changing the configuration is important. Literally hundreds of  papers, going back to the seminal work of \cite{kershaw} and perhaps beyond, use the Fisher Information as a measure of sensor performance.  In this context, parametrized families of  Fisher-Rao  metrics arise (\emph{e.g.} \cite{fishfam1,fishfam2}). Sensor management then 
becomes one of choosing an optimal metric (based on the value of the estimated parameter), from among a family of such, to permit the acquisition of the maximum amount of information about that parameter. 

As we have stated, the  focus, hopefully clarifying,  example  of this paper, is that of  estimating the location of a target using measurements from mobile sensors (\emph{c.f.} \cite{famdec1,famdec2}). The information content of the system depends both on the location of the target and on the spatial locations of the sensors, because the covariance of measurements is sensitive to the distances and angles made between the sensors and the target. As the sensors move in space, the associated likelihoods vary, as do the resulting  Fisher matrices, describing the information content of the system, for every possible sensor arrangement. It is this interaction between sensors and target that this paper sets out to elucidate in the context of information geometry.

The collection of all Riemannian metrics on a Riemannian manifold itself admits the structure of an infinite-dimensional Riemannian manifold \citep{gil1,gil2}. Of interest to us is only the subset of Riemannian metrics corresponding to Fisher informations of sensor configurations, and this allows us to restrict attention to a finite-dimensional sub-manifold of the manifold of metrics, called the \emph{sensor manifold} \citep{moran121,moran122}. In particular, a continuous adjustment of the sensor configuration, say by moving one of the individual sensors,  results in a continuous change in the associated Fisher metric and so a movement in the sensor manifold.

Though computationally difficult, the idea of regarding the Fisher metric as a measure of performance of a given sensor configuration and then understanding variation in sensor configuration in terms of the manifold of such metrics is powerful. It permits questions concerning optimal target trajectories, as discussed here, to minimize information passed to the sensors and, as will be discussed in a subsequent paper, optimal sensor trajectories to maximize information gleaned about the target.  In particular, we remark that the metric on the space of Riemannian metrics that appears naturally in a mathematical context in \cite{gil1,gil2}, also has a natural interpretation in a statistical context. 

Our aims here are to further  develop the information geometry view of sensor configuration begun in \cite{moran121,moran122}. While the systems discussed are simple and narrow in focus, already they point to concepts of information collection that appear to be new. Specifically, we set up the target location problem in an information geometric context and we show that the optimal (in a manner to be made precise in Sec. II) sensor trajectories, in a physical sense, are determined by solving the geodesic equations on the sensor manifold (Sec. III). Various properties of geodesics on this space are derived, and the mathematical machinery is demonstrated using concrete physical examples (Sec IV).


\section{The Information in Sensor Measurements}
\label{sec:inform-cont-sens}
\begin{figure*}[h]
  \begin{center}
\includegraphics[width=0.8\columnwidth]{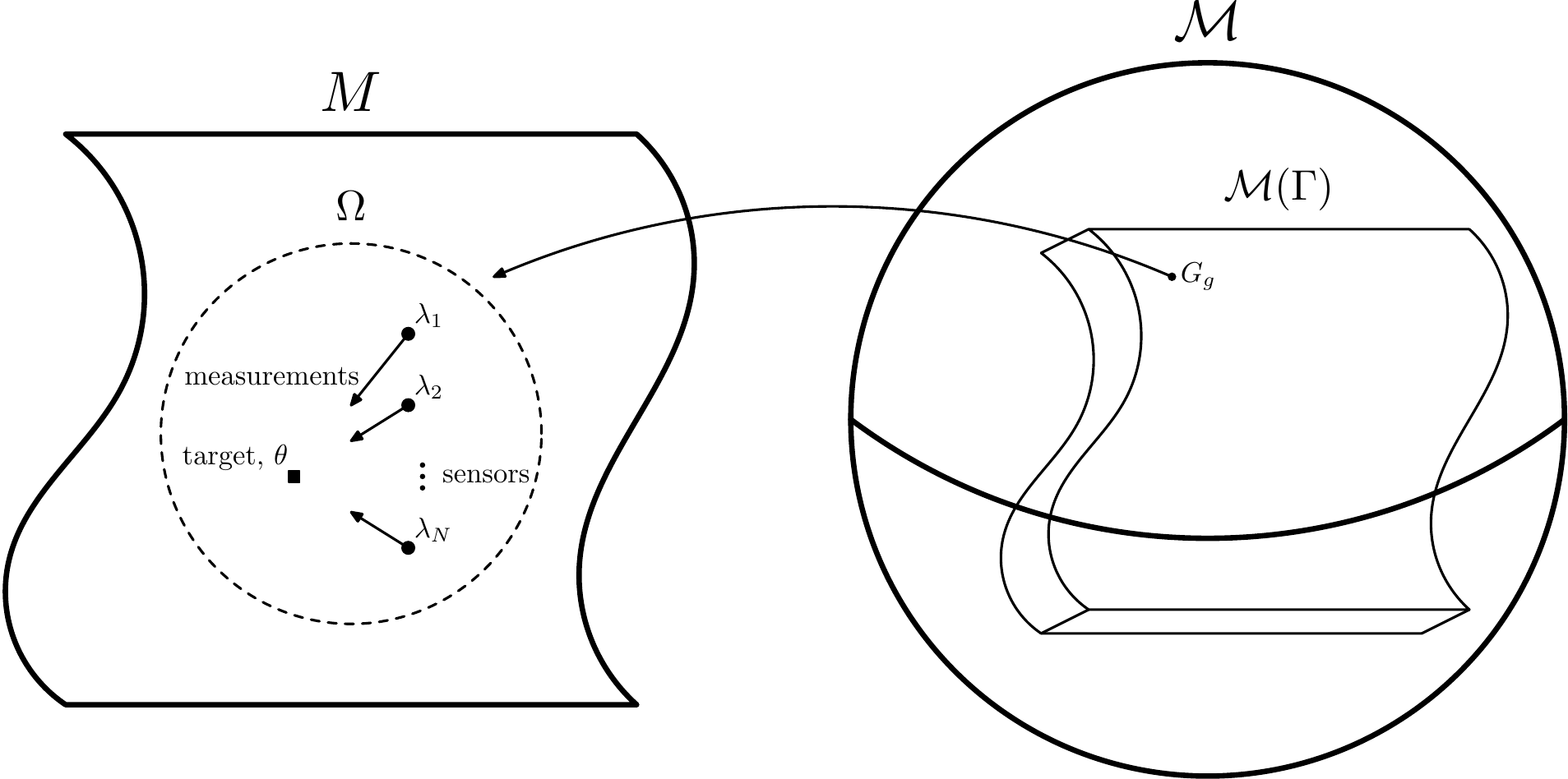}
\end{center}
\caption{Diagrammatic representation of the sensor model; sensors are at $\lambda_i\in M$ taking measurements of a target at $\bt\in M$.  A measure of distance between different sensor configurations, physically corresponding to change in information content is obtained through a suitable restriction of the metric $G_g$ \eqref{eq:gilmed} to the configuration manifold $\mathcal{M}(\Gamma) \subset \mathcal{M}$, the space of all Riemannian metric on $M$.  $\mathcal{M}$ is almost certainly not topologically spherical, it is merely drawn here as such for simplicity.  \label{geomsetup}}
\end{figure*}
In general, sensor measurements, as considered in this paper, can be formulated as follows.  Suppose we have, in a fixed  manifold $M$, a collection of $N$ sensors  located at $\lambda_i$ $(i = 1, \ldots , N )$. For instance the manifold may be $\mathbb R^3$ and location may just mean that in the usual sense in Euclidean space. The measurements from these sensors  are used to estimate the location of a target  $\bt$ also  in $M$ (see the left of Figure~\ref{geomsetup}).  Each sensor draws measurements ${x}_i$ from a distribution  with  probability density function (PDF) $p_i({x}_i|\bt)$. A measurement $\mathbf{x}=\{x_i\}_{i=1}^N$ is the collected set of individual measurement from each of the sensors with likelihood
\begin{equation}
  \label{eq:loglik}
p(\mathbf{x}|\bt) = \prod_{i=1}^N p_i({x}_i|\bt).
\end{equation}
Measurements here are assumed independent\footnote{While this assumption is probably not necessary, it allows one to define the aggregate likelihood \eqref{eq:loglik} as a simple product over the individual likelihoods, which renders the problem computationally more tractable.} between sensors and over time. 

Given a measurement $\mathbf{x}$ of a target at $\bt$, the likelihood that the same measurement could be obtained from a target at $\bt'$, $L(\bt,\bt')$, is given by the log odds expression
\[L(\bt,\bt') = \log\frac{p(\mathbf{x}|\bt)}{p(\mathbf{x}|\bt')},\]
   and the average over all measurements is, by definition, the Kullback-Leibler divergence \citep{kl}, $D(\bt||\bt')$:
   \begin{equation}
     \label{eq:6}
     D(\bt||\bt') = \mathbb{E}_{\mathbf{x}}\left[ L(\bt,\bt')\right] = \int p(\mathbf{x}|\bt) \log\frac{p(\mathbf{x}|\bt)}{p(\mathbf{x}|\bt')} \,d\mathbf{x}
   \end{equation}
 This would, ostensibly,  be a good measure on the information in the sensor measurements as it is non-negative and $D(\bt||\bt)=0$, but it lacks desirable features of a metric: it is not symmetric and does not satisfy the triangle inequality. We recall that the Kullback-Leibler divergence is related to mutual information, and refer the reader to Section 17.1 of \cite{coverthomas} for a discussion of this connection.   It is widely used as a measure of the difference in information available about the target between the locations $\bt$ and $\bt'$. 

In the limit as $\bt'\to\bt$ the first non-zero term in the series expansion for $D$ is second order, viz.
   \begin{equation}
     \label{eq:7}
     \lim_{\bt'\to\bt}D(\bt||\bt') = \lim_{\bt'\to\bt}(\bt-\bt')^T g (\bt-\bt') + O\left((\bt-\bt')^3\right),
   \end{equation}
where $g$ is an $n\times n$ symmetric matrix, and $n$ is the dimension of the manifold $M$.

This location-dependent matrix defines a metric over $M$, the \emph{Fisher Information Metric} \citep{fisher1,fisher2}. It can also be calculated, under mild conditions,  as the expectation of the tensor product of the gradients of the log-likelihood $\ell = \log p(\mathbf{x}|\bt)$ as
\begin{equation} \label{eq:fishdef}
g = \mathbb{E}_{\mathbf{x}|\bt} \left[ d_{\bt} \ell \otimes d_{\bt} \ell \right].
\end{equation}

Since Fisher Information is additive over independent measurements, the Fisher Information Metric provides a measure of the instantaneous change in information the sensors can obtain about the target. In this paper, we adopt a relatively simplistic view that the continuous case of measurements is a limit of measurements discretized over time.     Because sensor measurements depend on the relative locations of the sensors and target, this incremental change depends on the direction the target is moving; the Fisher metric \eqref{eq:fishdef} can naturally be expressed in coordinates that represent the sensor locations (see Sec. IV), but also depends on parameters that represent target location, which may be functions of time in a dynamical situation. Once the Fisher metric \eqref{eq:fishdef} has been evaluated, one can proceed to optimize it in an appropriate manner.





\subsection{D-Optimality}

Because the information of the sensor system described in Section~\ref{sec:inform-cont-sens} is a matrix-valued function it is not obvious what it means to maximize the `total information' with respect to the sensor parameters. We require a definition of `optimal' in an information theoretic context. Several different optimization criteria exist (\emph{e.g.} \cite{dopt1,dopt2}),  defined by constructing scalar invariants from the matrix entries of $g$, and maximizing those functions in the usual way. 

We adopt the notion of \emph{D-optimality} in this paper; we consider the maximization of  the determinant of \eqref{eq:fishdef}. Equivalently, D-optimality maximizes the differential Shannon entropy of the system with respect to the sensor parameters \citep{seb97}, and  minimizes  the volume of the elliptical confidence regions for the sensors estimate of the location of the target $\bt$ \citep{dopt3}.

A complication in applying  D-optimality (or any other) criterion to this problem is that  the sensor locations and distributions are not fixed. Conventionally,  measurements are  drawn from sensors with \emph{fixed} properties, with a view to estimating a parameter $\bt$. Permitting sensors to move throughout $M$ produces an infinite family of sensor configurations, and hence Fisher-Rao metrics \eqref{eq:fishdef}, parametrized by the locations of the sensors. One aim of this structure is to move sensors to locations that serve to maximize  information content, given some prior distribution for a particular $\bt \in M$. This necessitates a tool to measure the difference between information provided by members of a family of Fisher-Rao metrics; this is explored in Section~\ref{sec:conf-manif}.


\subsection{Geodesics on the Sensor Manifold}
\label{sec:geod-sens-mani}


We now consider the  case where the target is in motion, so that $\bt$ varies along a path $\gamma(t)\subset M$. The instantaneous information gain by the sensor(s) at time $t$ is then $ g\left(\gamma'(t),\gamma'(t)\right)$, where $g$ is the Fisher Information Metric \eqref{eq:fishdef}. This observation is based on the assumption that  the measurements are all independent.   The total information $I$ gained along $\gamma$  is 
\begin{equation}
  \label{eq:3}
I(T) =  \int_0^T g(\gamma'(t),\gamma'(t)) \,dt,
\end{equation}
which is the equivalent of the energy functional in differential geometry [e.g. Chapter 9 of \cite{docarmo}], and this has the same extremal paths as $l_{g}(\gamma)$, the arc-length of the path $\gamma$,
\begin{equation}
  \label{eq:2}
  l_{g}(\gamma) = \int_0^T \sqrt{g\left(\gamma'(t),\gamma'(t)\right)}\,dt.
\end{equation}
Paths with extremal values of this length are \emph{geodesics} and these can be interpreted as the evasive action that can be taken by the target to minimize amount of information it gives to the sensors.

\subsection{Kinematic Conditions on Information}
\label{sec:kinconst}\stodo{Please read and critique; Attempted AS 03/10}
While the curves that  are extrema of the Information functional and of  arc-length are the same as sets,  a geodesic  only minimizes the  information  functional if traversed at speed 
\begin{equation}
\label{eq:speed}
dl_{g}/dt = +\sqrt{g(\gamma'(t),\gamma'(t))}.
\end{equation}
In differential geometric terms, this is equivalent to requiring the arc-length parametrization of the geodesic to fulfill the energy condition. 
In order to minimize information about its
location, the target should move along a geodesic of $g$ at exactly the speed
$dl_{g}/dt$ \eqref{eq:speed}. This direct kinematic condition on information is unusual and difficult to reconcile with our current view of information theory. 

While aspects of this speed constraint are still unclear to us, 
an analogy that may be useful is to regard the target as moving though a ``tensorial information fluid''. Now moving slower relative to the fluid will result in ``pressure'' building behind, requiring information (energy) to be expended to maintain the slower speed. Moving faster also requires more information to push through the slower moving fluid.
In the fluid dynamics analogy,  the energy expended in moving though a fluid is proportional to the square of the difference in speed between the fluid and the object. The local energy is proportional to the difference between actual speed and the speed desired by the geodesic; that is, the speed that minimizes the energy functional.  Pursuing the fluid dynamics analogy, the density of the fluid mediates the relationship between the energy and the relative speed. 
\begin{equation*}
E\propto g(\delta \mathbf v,\delta \mathbf v)
\end{equation*}
In particular, the scalar curvature, which depends on $G$,  influences the energy and hence the information flow. We will explore this issue further in a future publication. 

\section{The Information of Sensor Configurations}
\label{sec:conf-manif}

A sensor configuration is a set $\Gamma=\{\lambda_i\}_{i=1}^N$ of sensor parameters. The Fisher-Rao metric $g$ can be viewed as a function of $\Gamma$ as well as $\theta$, the location of the target. To calculate the likelihood that a measurement came from one configuration $\Gamma_0$ over another $\Gamma_1$ requires the calculation of $p(x|\Gamma_0,\theta)/p(x|\Gamma_1,\theta)$, which is difficult as the value of $\theta$ is not known exactly. Measurements can be used to construct an estimate $\hat\theta$, however, the distribution of this estimate is hard to quantify and even harder to calculate.  Instead, here,   the maximum entropy distribution is used. This is normally distributed with mean $\hat\theta$, and covariance $g^{-1}(\hat\theta)$, the inverse of the Fisher information metric at the estimated target location.


The information gain due to the sensor configuration $\Gamma$ is now $D(p(x|\Gamma,\hat{\theta})||1)$ because there was no prior information about the location (the uniform distribution\footnote{Note that the uniform distribution is, in general, an improper prior in the Bayesian sense unless the manifold $M$ is of finite volume. It may be necessary, therefore, to restrict attention to some (compact) submanifold $\Omega \subset M$ for $D(p(x|\Gamma,\hat{\theta})||1)$ to be well-defined; see also the discussion below equation \eqref{eq:8}.}) before the sensors were configured  compared with the maximum entropy distribution $p$ after. Evaluating this gives
\begin{equation}
  \label{eq:1}
  D(p||1) = \log\left((2\pi e)^n\det g^{-1}(\Gamma,\hat{\theta})\right)
\end{equation}
The Fisher Information metric $G$ for this divergence can be calculated from
\begin{equation}
  \label{eq:8}
  G(h,k) = \sE\left[d_g^2D(p||1)\right] = \int_M tr(g^{-1}hg^{-1}k)\vol(g)\,d\mu
\end{equation}
where $h$ and $k$ are tangent vectors to the space of metrics. The integral defining \eqref{eq:8} may not converge for non-compact $M$, so restriction to a compact submanifold $\Omega$ of $M$ is assumed throughout as necessary (\emph{c.f.} Figure~\ref{geomsetup}).

\subsection{The Manifold of Riemannian Metrics}
\label{sec:manif-riem-metr}

The set of all Riemannian metrics over a manifold $M$ can itself be imbued with the structure of an infinite-dimensional Riemannian manifold \citep{gil1,gil2}, which we call $\mathcal{M}$. Points of $\mathcal{M}$ are Riemmanian metrics on $M$; i.e. each point $G  \in \mathcal{M}$ bijectively corresponds to a positive-definite, symmetric $(0,2)$-tensor in the space $S^{2}_{+}T^{\star}M$.  Under reasonable assumptions, an $L^{2}$ metric on $\mathcal{M}$ \citep{clarkephd,gil1} may be defined as:
\begin{equation} \label{eq:gilmed}
G(h,k) = \int_M tr(g^{-1}hg^{-1}k)\vol(g)\,d\mu,
\end{equation}
which should be compared to \eqref{eq:8}. 


It should be noted that the points of the manifold $\mathcal{M}$ comprise \emph{all} of the metrics that can be put on $M$, most of which are irrelevant for our physical sensor management problem.  We restrict  consideration to a sub-manifold of $\mathcal{M}$ consisting only of those Riemannian metrics that are members of the family of Fisher information matrices \eqref{eq:fishdef} corresponding to feasible sensor configurations. This particular sub-manifold is called the `sensor' or `configuration' manifold \citep{moran121,moran122} and is denoted by $\mathcal{M}({\Gamma})$, where now the objects $h$ and $k$ are now elements of the now finite dimensional tangent space $T \mathcal{M}({\Gamma})$. The dimension of $\mathcal{M}({\Gamma})$ is $N \times \dim(M)$ since each point of $\mathcal{M}({\Gamma})$ is uniquely described by the locations of the $N$-sensors, each of which require $\dim(M)$ numbers to denote their coordinates. A visual description of these spaces is given in Figure~\ref{geomsetup}. For all cases considered in this paper, the integral defined in \eqref{eq:gilmed} is well-defined and converges (see however the discussion in \cite{famdec1}). 



For the purposes of computation, it is convenient to have an expression for the metric tensor components of \eqref{eq:gilmed} in some local coordinate system. In particular, in a given coordinate basis $z^{i}$ over $\mathcal{M}({\Gamma})$ (not to be confused with the coordinates on $\Omega$; see Sec. IV), the metric \eqref{eq:gilmed} reads
\begin{equation} \label{eq:metten}
G(h,k) = \int_{\Omega}  g^{nk}  g^{\ell m} h_{mn}  k_{\ell k}  \text{vol}(g).
\end{equation}
where $h$ and $k$ are tangents vectors in $T \mathcal{M}({\Gamma})$ given in coordinates by
\begin{equation}
  \label{eq:9}
  T \mathcal{M}(\Gamma) = \text{span}\left\{ \frac{\partial}{\partial z^i} g_{mn}\right\}_{i=1}^{\dim \mathcal{M}({\Gamma})}.
\end{equation}
From the explicit construction \eqref{eq:metten}, all curvature quantities of $\mathcal{M}({\Gamma})$, such as the Riemann tensor and Christoffel symbols, can be computed. 



\subsection{D-Optimal Configurations}

\begin{figure*}[h]
\includegraphics[width=\columnwidth]{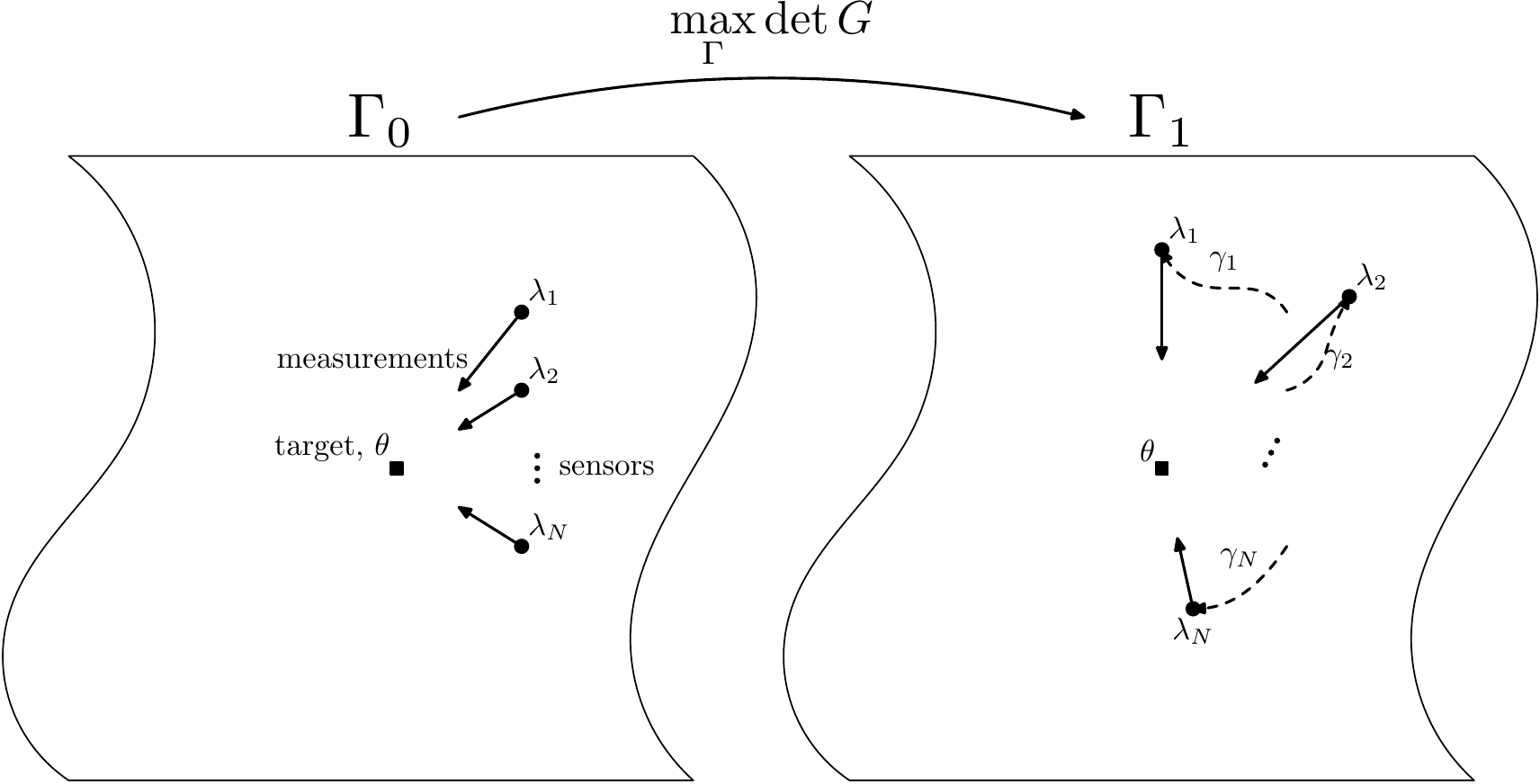}
\justifying
\caption{Graphical illustration of D-optimal sensor dynamics; a sensor configuration $\Gamma_0$ evolves to a new configuration $\Gamma_1$ by moving the $N$ sensors in $\Omega$-space to new positions that  are determined by maximizing the determinant of the metric ${G}$, given by equation \eqref{eq:metten}, on the sensor manifold $\mathcal{M}({\Gamma})$. Each sensor $\lambda^{i}$ traverses a path $\gamma_{i}$ through $\Omega$-space to end up in the appropriate positions constituting $\Gamma_1$. As shown in Section \ref{sec:geod-conf-manif}, the paths $\gamma_{i}$ are entropy-minimizing if they are geodesic on the sensor manifold $\mathcal{M}({\Gamma})$. Note that the target is shown as stationary in this particular illustration. \label{evolvingsensor}
}
\end{figure*}

D-optimality in the context of the sensor manifold described above is discussed in this section. Suppose that the sensors are arranged in some arbitrary configuration $\Gamma_{0}$. The sensors now move in anticipation of target behaviour; a prior distribution is adopted to localize a target position $\bt$. The sensors move continuously to a new configuration $\Gamma_{1}$, where $\Gamma_{1}$ is determined by maximizing the determinant of $G$, i.e. $\Gamma_{1}$ corresponds to the sensor locations for which $\det({G})$, computed from \eqref{eq:metten}, is maximized. The physical situation is depicted graphically in Figure \ref{evolvingsensor}. This process can naturally be extended to the case where real measurement data is used. In particular, as measurements are drawn, a (continuously updated) posterior distribution for $\bt$ becomes available, and this can be used to update the Fisher metric (and hence the metric $G$) to define a sequence $\Gamma_{t}$ of optimal configurations; see Sec. V.


\subsection{Geodesics for the Configuration Manifold}
\label{sec:geod-conf-manif}

While D-optimality allows us to determine \emph{where} the sensors should move given some prior, it provides us with no guidance on which \emph{path} the sensors should traverse, as they move through $\Omega$, to reach their new, optimized positions.

A path $\Upsilon(t)$ from one configuration $\Gamma_0$ to another $\Gamma_1$ is  a set of paths $\Gamma(t) = \left\{\gamma_i(t)\right\}_{i=1}^N$ for each sensor $S^i$ from location $\lambda_i^0$ to $\lambda_i^1$. Varying the sensor locations is equivalent to varying the metric $g(t) = g(\Gamma(t),\hat{\theta}(t))$ and the estimate of the target location $\hat{\theta}$. The information gain along $\Upsilon$ is then
\begin{equation}
  \label{eq:10}
  \int_\Upsilon G_{g(t)}\left(g'(t),g'(t)\right)\, dt,
\end{equation} 
and the extremal paths are again the  geodesics of the metric $G$. Also the speed constraint observed earlier in Section \ref{sec:inform-cont-sens}.\ref{sec:kinconst} for the sensor geodesics is in place here and given by
\begin{equation}
  \label{eq:11}
  \frac{dl_G}{dt} = \sqrt{G_{g(t)}\left(g'(t),g'(t)\right)}.
\end{equation}
Again, this leads to the conclusion that there are kinematic constraints on the rate of change of sensor parameters that lead to the collection of the maximum amount of information.

\section{Configuration for Bearings-only Sensors}

To illustrate the mathematical machinery developed in the previous sections consider first the  configuration metric for two bearings-only sensors. The physical space $\Omega$, where the sensors and the target reside, is chosen to be the square $[-10,10] \times [-10,10] \subset \mathbb{R}^{2}$. Effectively, we assume an uninformative (uniform) prior over the square $\Omega$.

The goal is to estimate a position $\bt$ from bearings-only measurements taken by the sensors, as in previous work \citep{moran121,moran122}.  We assume that measurements are drawn from a Von Mises distribution,
\begin{equation} \label{eq:mises}
  M_n \sim p_n(\cdot|\theta)={\frac {e^{\kappa \cos\left[\cdot -\arg\left(\theta-{\lambda}_{n} \right) \right]}}{2\pi I_{0}(\kappa )}},
\end{equation}
where $\kappa$ is the concentration parameter and $I_r$ is the $r$th modifed Bessel function of the first kind,   \citep{misesvar}, and ${\lambda}_{n} = (x_{n},y_{n})$ is the location of the $n$-th sensor in Cartesian coordinates. Note that, in reality, the parameter $\kappa$ will depend on location, since the signal-to-noise ratio will decrease as the sensors move farther away from the target. This is beyond the scope of this paper and will be addressed in future work.\bill{I think Xuezhi would say that since we're measuring angle the distance doesn't matter. This needs to be clarified}

For the choice \eqref{eq:mises}, the Fisher metric \eqref{eq:fishdef} can be computed easily, and has components
\begin{multline} \label{eq:bearingsfish}
{g} = \kappa \left(1-\frac{I_2(\kappa)}{2I_0(\kappa)}\right)\\ \sum_{i=1}^N \frac1{(x-x_i)^2+{(y-y_i)}^2}
  \begin{pmatrix}
    (y-y_i)^2 & -(x-x_i)(y-y_i)\\
-(x-x_i)(y-y_i) & (x-x_i)^2
  \end{pmatrix}.
\end{multline}

\subsection{Target Geodesics}
\label{sec:visu-targ-geod}
A geodesic, $\bg(t)$, starting at $\mathbf{p}$ with initial direction $\mathbf{v}$ for a manifold with metric $g$ is the solution of the coupled second-order initial value problem for the components $\gamma^i(t)$:
\begin{equation}
\label{eq:geoivp}
 \frac{d^2\gamma^i}{dt^2} = \Gamma^i_{jk}\frac{d\gamma^j}{dt}\frac{d\gamma^k}{dt},\quad \bg(0)=\mathbf{p},\
 \bg'(0)=\mathbf{v},
 \end{equation}
where $\Gamma^i_{jk}$ are the Christoffel symbols for the metric $g$.

\begin{figure}[h]
  \centering
  \includegraphics[width=0.9\columnwidth]{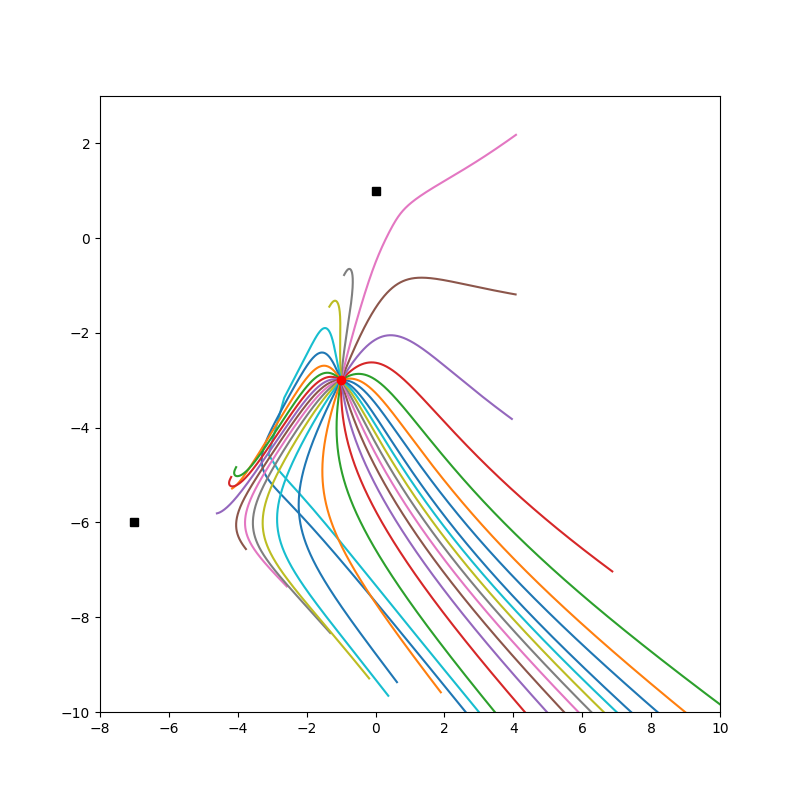}
  \caption{Solutions to the geodesic equation on $\Omega = [-10,10]\times [-10,10]$ for a target starting at $(-1,-3)$ and sensors  at $(-7,-6)$ and $(0,1)$. The differing paths correspond to the initial direction vector $(\cos\theta,\sin\theta)$  of the target, varying as $\theta$ varies from 0 to $2\pi$ radians in steps of 0.25 radians.}
  \label{fig:geodesic_eqn}
\end{figure}

Figure \ref{fig:geodesic_eqn} shows directly integrated solutions to the geodesic equation \eqref{eq:geoivp} a target at $(-1,-3)$ and sensors  at $(-7,-6)$ and $(0,1)$. The differing paths correspond to the initial direction vector $(\cos\phi,\sin\phi)$ of the target, varying as $\phi$ varies from 0 to $2\pi$ radians in steps of $0.25$ radians. 

An alternative way to numerically compute the geodesics connecting two points on the manifold $M$ is using the Fast Marching Method~(FMM) \cite{sethian1996fast, sethian1999fast,tsitsiklis1995efficient}. Since the Fisher-Rao Information Metric is a Riemannian metric on the Riemannian manifold $M$, 
one can show that the geodesic distance map $u$, the geodesic distance from the initial point $\mathbf{p}$ to a point $\bt$, satisfies the Eikonal equation 
\begin{equation}\label{eq: eikonal}
|| \nabla u(\bt) ||_{g_{\bt}^{-1}} = 1
\end{equation}
with initial condition $u(\mathbf{p}) = 0$. By using a mesh over the parameter space, the (isotropic or weakly anisotropic) Eikonal equation (\ref{eq: eikonal}), can be solved numerically by Fast Marching. The geodesic is then extracted by integrating numerically the ordinary differential equation
\begin{equation}
\frac{d \bg(t)}{dt} = -\eta_t g_{\bt(t)}^{-1} \nabla u(\bt(t)),
\end{equation}
where $\eta_t > 0$ is the mesh size. 
The computational complexity of FMM is $O(N \log N)$, where $N$ is the total number of mesh grid points. For Eikonal equations with strong anisotropy, a generalized version of FMM, Ordered Upwind Method~(OUM)~ \cite{sethian2003ordered} is preferred.

\begin{figure}[h]
  \centering
  \includegraphics[width=0.9\columnwidth]{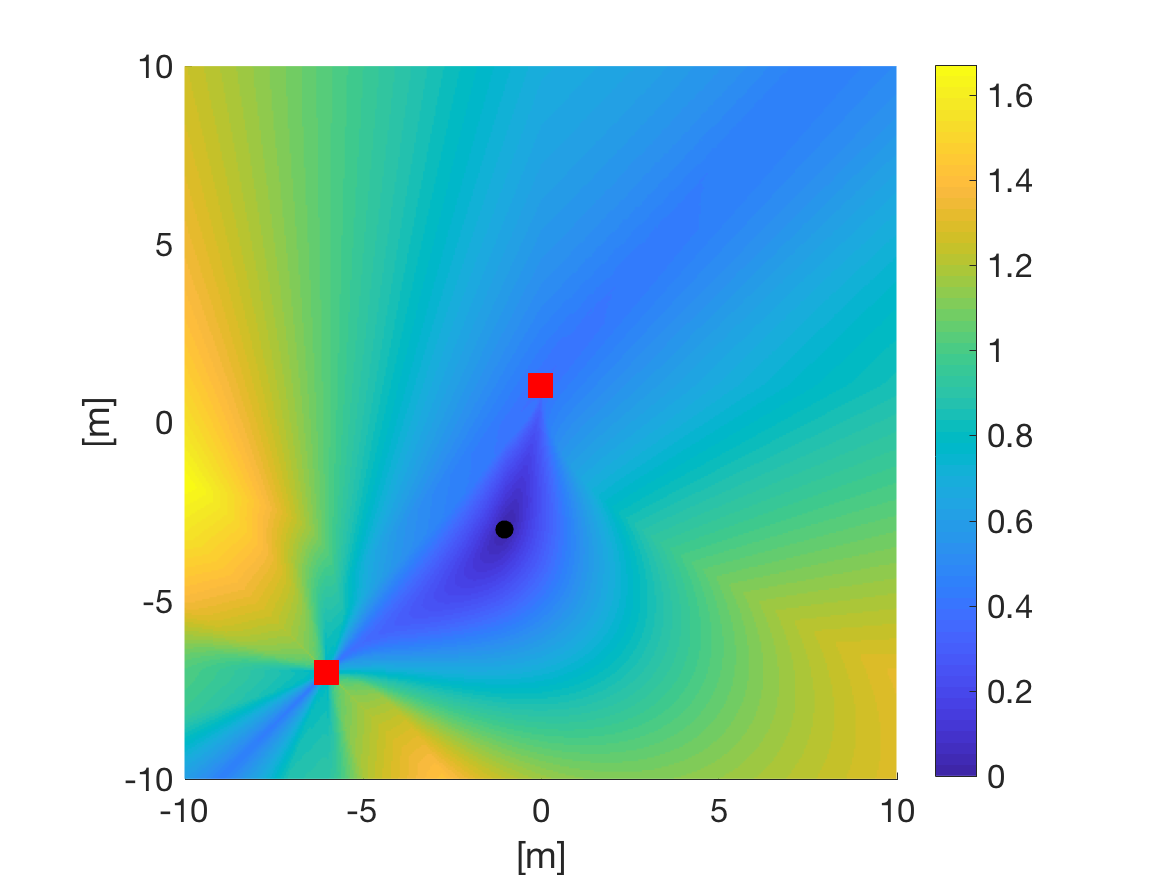}
  \caption{Geodesic distance on $\Omega = [-10,10]\times [-10,10]$ from the point $(-1,-3)$ with  sensors  at $(-7,-6)$ and $(0,1)$. The distance was calculated using a Fast Marching formulation of the geodesic equation. The fact that the geodesics follow the gradient of this distance allows comparison with Figure~\ref{fig:geodesic_eqn} }
  
  \label{fig:geodesic_fm}
\end{figure}

Compare Figure~\ref{fig:geodesic_eqn} with Figure~\ref{fig:geodesic_fm} which uses a Fast Marching algorithm to calculate the geodesic distance from the same point. 

\begin{figure}[h]
  \centering
  \includegraphics[width=0.9\columnwidth]{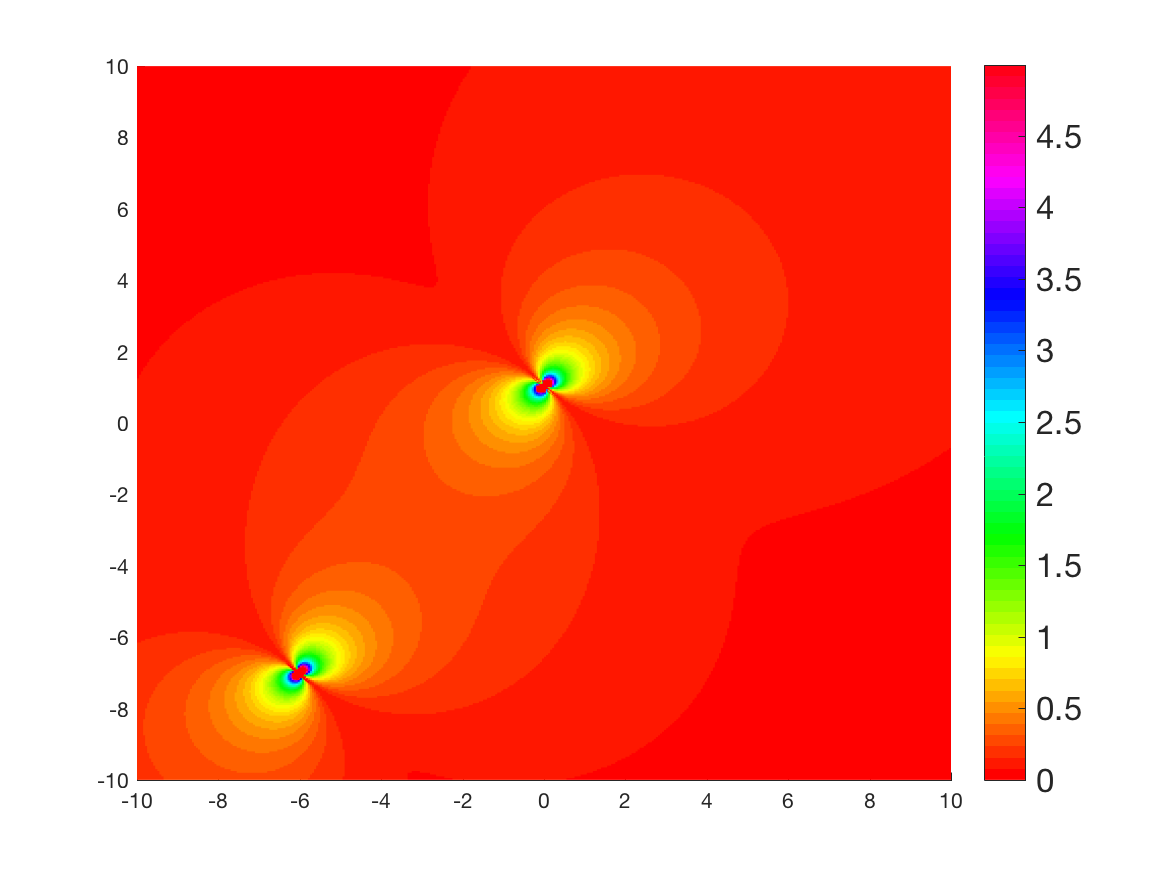}
  \caption{Geodesic speed for each point in $\Omega = [-10,10]\times [-10,10]$ for targets departing in  direction of the vector (1,-1)}
  \label{fig:geodesic_speed}
\end{figure}

Figure~\ref{fig:geodesic_speed} shows the speed required along a geodesic travelling in the direction of the vector $(1,-1)$. \stodo{Again critique required}In the case of these bearing-only sensors the trade-off over speed is between time-on-target and rate of change of relative angle between the sensors and the target. Travelling slowly means more time for the sensors to integrate measurements but the change in angle is slower,  resulting in more accurate position estimates. Conversely, faster movement than the geodesic speed results in larger change in angle measurements but less time for measurements again resulting in more accurate measurements. Only at the geodesic speed is the balance  reached and the minimum information criterion achieved.

\subsection{Configuration Metric Calculations}
\label{sec:conf-metr-calul}

The coordinates $\boldsymbol{z}$ on $S_{\Omega}$ are $\{x_{1},y_{1},x_{2},y_{2}\}$. The sensor management problem amounts to, given some initial configuration, identifying a choice of $\{x_{1},y_{1},x_{2},y_{2}\}$ for which the determinant of \eqref{eq:metten} is maximized, and traversing geodesics $\gamma^{i}$ in $S_\Omega$,  starting at the initial locations and ending at the D-optimal locations; see Figure~\ref{evolvingsensor}. We assume that the target location is given by an uninformative prior distribution; that is,  $\mathbb{P}(\bt \in A)  = \text{vol}(A)/\text{vol}(\Omega)$ for all $A \subset \Omega$. 

\begin{figure}[h]
\includegraphics[width=0.9\columnwidth]{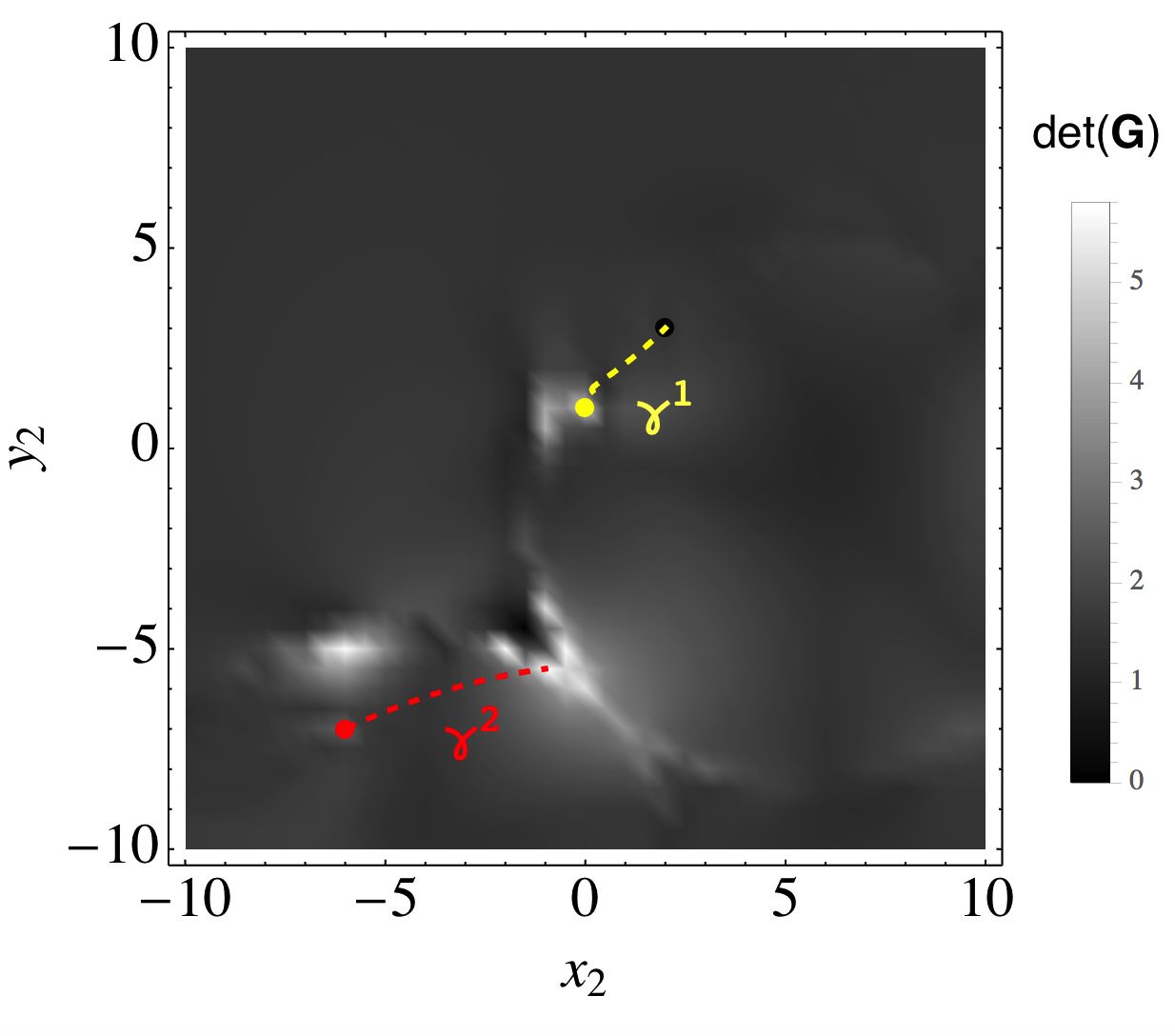}
\justifying
\caption{Contour plot of $\det(\boldsymbol{G})$, for $\boldsymbol{G}$ given by \eqref{eq:metten}, for the case where $S^{1}$ moves from $(0,1)$ (yellow dot) to $(2,3)$ (black dot). The second sensor $S^{2}$ begins at the point $(-6,-7)$ (red dot). Brighter shades indicate a greater value of $\det(\boldsymbol{G})$; $\det(\boldsymbol{G})$ is maximum at $(-1,-5.5)$. The geodesic path linking the initial and final positions of $S^{1}$ is shown by the dashed yellow curve, while the dashed red curve shows the geodesic path linking $(-6,-7)$, the initial position of $S^{2}$, to the D-optimal location $(-1,-5.5)$. \label{detg1}
}
\end{figure}

To make the problem tractable, we consider a simple case where one of the sensor trajectories is fixed; that is,  we consider a 2-dimensional submanifold of $\mathcal{M}(\Gamma)$ parametrized by the coordinates $(x_{2},y_{2})$ of $S_{2}$ only. Figure \ref{detg1} shows a contour plot of $\det({G})$, as a function of $(x_{2},y_{2})$, for the case where $S_{1}$ moves from $(0,1)$ (yellow dot) to $(2,3)$ (black dot). The second sensor $S_{2}$ begins at the point $(-6,-7)$ (red dot). Figure~\ref{detg1} demonstrates that, provided $S_{1}$ moves from $(0,1)$ to $(2,3)$, $\det({G})$ is maximized at the point $(-1,-5.5)$, implying that $S_{2}$ moving to $(-1,-5.5)$ is the D-optimal choice. The geodesic path $\gamma_{1}$ beginning at $(0,1)$ and ending at $(2,3)$ is shown by the dashed yellow curve; this is the information-maximizing path of $S_{1}$ through $\Omega$. Similarly for $S_{2}$, the geodesic path $\gamma^{2}$ beginning at $(1,-3)$ and ending at $(-1,-5.5)$ is shown by the dotted red curve. 

\subsection{Visualizing Configuration Geodesics}
\label{sec:visu-config-geod}

\begin{figure}[h]
  \centering
  \includegraphics[width=0.9\columnwidth]{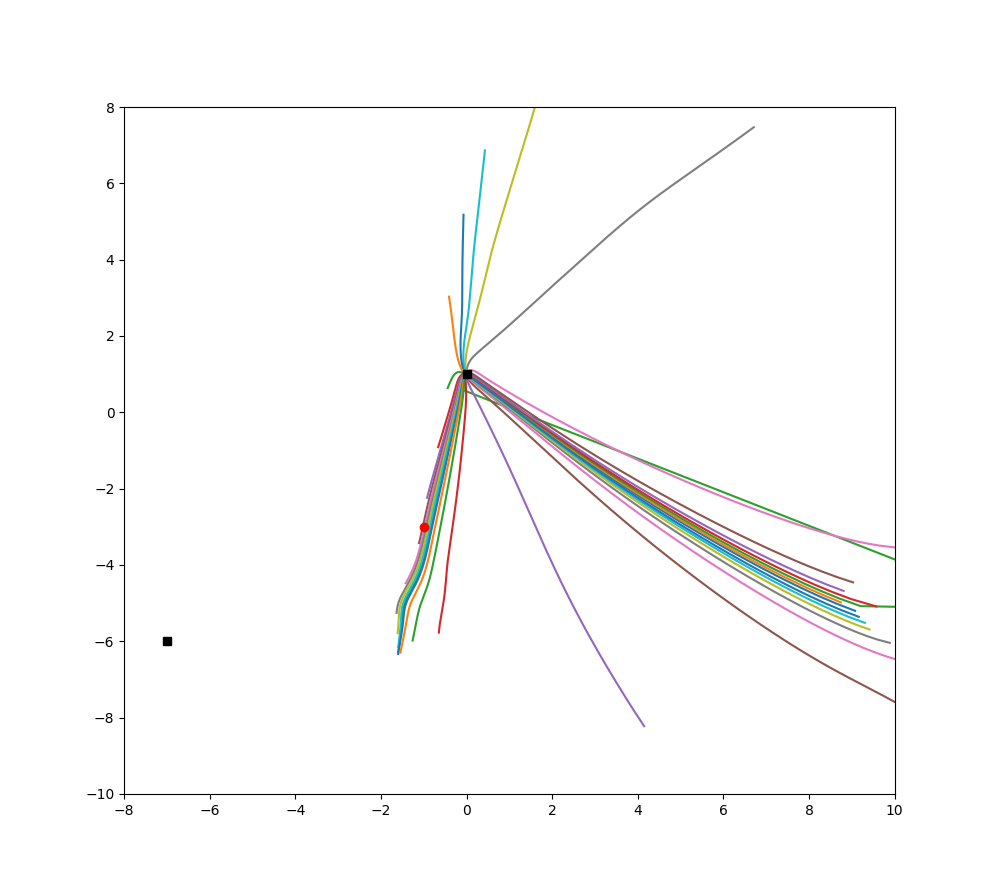}
  \caption{Solutions to the geodesic equation on $\Omega = [-10,10]\times [-10,10]$ for a sensor starting at $(0,1)$ with the other sensor stationary at $(-7,-6)$ and the target at (-1,-3). The differing paths correspond to the initial direction vector $(\cos\phi,\sin\phi)$ of the target, varying as $\phi$ varies from 0 to $2\pi$ radians in steps of 0.25 radians.}
  \label{fig:cgeodesic_eqn}
\end{figure}

\begin{figure}[h]
  \centering
  \includegraphics[width=0.9\columnwidth]{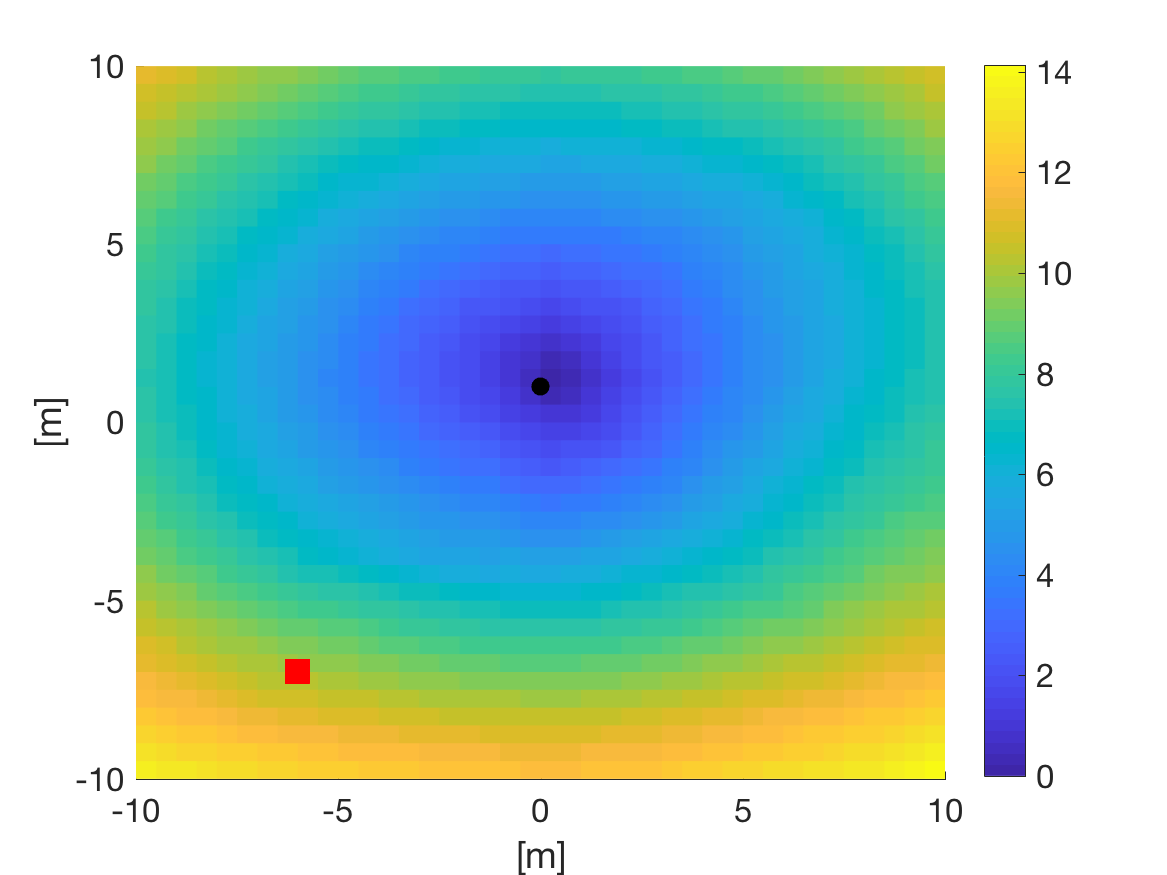}
  \caption{Geodesic distance on $\Omega = [-10,10]\times [-10,10]$ for a sensor starting at $(0,1)$ with the other sensor stationary at $(-7,-6)$ and the using an uninformative prior for the target. The distance was calculated using a Fast Marching formulation of the geodesic equation. The fact that the geodesics follow the gradient of this distrance allows comparison with Figure~\ref{fig:cgeodesic_eqn}}
  \label{fig:cgeodesic_fm}
\end{figure}

\begin{figure}[h]
  \centering
  \includegraphics[width=0.9\columnwidth]{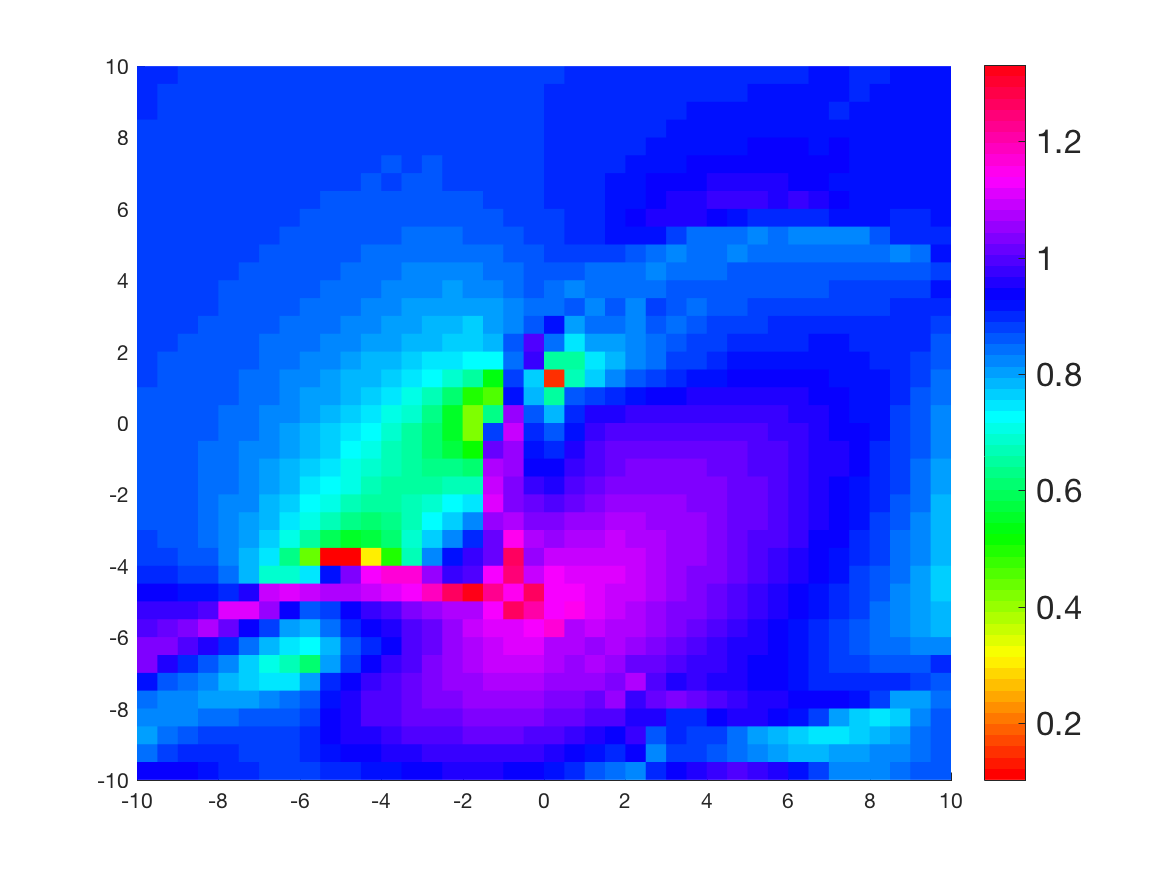}
  \caption{Geodesic speed sensor moving along a geodesic in direction (1,-1) at each point in $\Omega = [-10,10]\times [-10,10]$. The other sensor is stationary at $(-7,-6)$, and an uninformative prior is used for the target.}
  \label{fig:cgeodesic_speed}
\end{figure}

Figure \ref{fig:cgeodesic_eqn} shows solutions to the geodesic equation on $\Omega = [-10,10]\times [-10,10]$ for a sensor starting at $(0,1)$ with the other sensor stationary at $(-7,-6)$ and the target at $(-1,-3)$. The differing paths correspond to $\phi$ varying through $[0,2\pi]$ in steps of $0.25$ radians   in the  target initial direction vector $(\cos\phi,\sin\phi)$. It is interesting to note the concentration in direction of the geodesics in spite of the even distribution of initial directions. Both groups improve information, in a way that is well understood for bearing-only sensors, by increasing the rate of change of bearing. One group acheives this by closing with the target, the other by moving at right angles. This should be compared with  Figure~\ref{fig:cgeodesic_fm}, which uses a Fast Marching algorithm \citep{SethianJournal1998,peyre2010geodesic} to calculate the geodesic distance from the same point. 
Figure~\ref{fig:cgeodesic_speed} shows the speed required along a geodesic travelling  through each point in the direction of the vector (1,1). 









\section{Discussion}
We consider the problem of sensor management from an information-geometric standpoint \citep{amari1,amari2}. A physical space houses a target, with an unknown position, and a collection of mobile sensors, each of which takes measurements with the aim of gaining information about target location \citep{moran121,moran122,optimpaper}. The measurement process is parametrized by  the relative positions of the sensors. For example, if one considers an array of sensors that take bearings-only measurements to the target \eqref{eq:mises}, the amount of information that can be extracted regarding  target location clearly depends on the angles between the sensors. In general, we illustrate that in order to optimize the amount of information the sensors can obtain about the target, the sensors should move to positions which maximize the norm of the volume form (`D-optimality') on a particular manifold imbued with a metric \eqref{eq:metten} which measures the distance (information content difference) between Fisher matrices \citep{seb97,dopt3}. We also show that, if the sensors move along geodesics [with respect to \eqref{eq:metten}]  to reach the optimal configuration, the amount of information that they \emph{give away} to the target is minimized. This paves the way for (future) discussions about game-theoretic scenarios where both the target and the sensors are competitively trying to acquire information about one another from stochastic measurements; see e.g. \cite{game1,game2} for a discussion on such games. Differential games along these lines will be addressed in forthcoming work.

We hope that this work may eventually have realistic applications to signal processing problems involving parameter estimation using sensors. We have demonstrated that there is a theoretical way of choosing sensor positions, velocities, and possibly other parameters in an optimal manner  so  that the maximum amount of useful data can be harvested from a sequence of measurements taken by the sensors. For example, with sensors that take continuous or discrete measurements, this potentially allows one to design a system that  minimizes the expected amount of time taken to localize (with some given precision) the position of a target. If the sensors move along paths that are geodesic with respect to \eqref{eq:metten}, then the target, in some sense, learns the least about its trackers. This allows the sensors to prevent either intentional or unintentional evasive manoeuvres; a unique aspect of information-geometric considerations. Ultimately,  these ideas may lead to improvements on search or tracking strategies available in the literature [e.g. \cite{track1,track2}].  Though we have only considered simple sensor models in this paper, the machinery can, in principle, be adopted to systems of arbitrary complexity. It would certainly be worth testing the theoretical ideas presented in this paper experimentally using various sensor setups.

\section*{Acknowledgements}
This work was supported in part by the US Air Force Office of Scientific Research (AFOSR) Under Grant No. FA9550-12-1-0418. 
Simon Williams acknowledges an Outside Studies Program grant from Flinders University.
All the authors declare that they have no further conflict of interest.




\bibliographystyle{spbasic}
\bibliography{paperbib}
\end{document}